# Probing the Dynamics of Spontaneous Quantum Vortices in Polariton Superfluids

K. G. Lagoudakis[1], F. Manni[1], B. Pietka[1], M. Wouters[2], T. C. H. Liew[2], V. Savona[2], A. V. Kavokin[3], R. André[4] and B. Deveaud-Plédran[1]

[1]*Institute of Condensed Matter Physics (ICMP), Ecole Polytechnique Fédérale de Lausanne, CH-1015, Lausanne, Switzerland*
[2]*Institute of Theoretical Physics (ITP), Ecole Polytechnique Fédérale de Lausanne, CH-1015, Lausanne, Switzerland*
[3]*Physics and Astronomy School, University of Southampton, Highfield, Southampton, SO171BJ, UK*
[4]*Institut Néel, CNRS, Grenoble, France*

The experimental investigation of spontaneously created vortices is of utmost importance for the understanding of quantum phase transitions towards a superfluid phase, especially for two dimensional systems that are expected to be governed by the Berezinski-Kosterlitz-Thouless physics. By means of time resolved near-field interferometry we track the path of such vortices, created at random locations in an exciton-polariton condensate under pulsed non-resonant excitation, to their final pinning positions imposed by the stationary disorder. We formulate a theoretical model that successfully reproduces the experimental observations.

The spontaneous formation and kinetics of quantum vortices in two dimensional bosonic quantum fluids is a subject of fundamental interest. The vortices are crucially involved in the appearance of superfluidity [1]. For an interacting two dimensional Bose gas, a superfluid phase can be formed below a critical temperature as a result of the phase transition that is associated with binding of pairs of vortices with opposite topological charges: the Berezinski-Kosterlitz-Thouless (BKT) transition [2]. It has been argued though that for realistic finite size 2D ultracold atomic gases or exciton-polariton systems, the BKT transition is not a valid picture and a finite size Bose-Einstein condensate (BEC) may be formed at finite temperatures instead of a superfluid [3]. Out of equilibrium, a pathway to the creation of spontaneous quantum vortices is the Kibble-Zurek mechanism [4],[5]. When an interacting Bose gas is brought rapidly from the incoherent to coherent phase, the order parameter forms locally, and a domain structure appears. The merging of the order parameter domains having different phases leads to formation of topological defects, as those observed recently for atomic condensates [6].

In atomic condensates, up to now the spontaneously created vortices have been only monitored by destructive methods: the condensate is allowed to expand, then the vortices are recorded by the acquisition of single snapshots [2],[6]. Although this method is elegant and efficient, it does not allow for the in depth investigation of the dynamics of neither the formation nor the propagation of the vortices.

In this Letter we study, the dynamics of spontaneously created vortices in a two dimensional non-resonantly generated quantum fluid of exciton-polaritons in a semiconductor microcavity [7]. So far, vorticity in polariton quantum fluids has only been investigated for the cases of pinned [8] or resonantly imprinted vortices [9],[10]. The dynamical behavior of *spontaneously created vortices* under *non-resonant* excitation remained unknown. Although at first sight the non-equilibrium character of polariton systems (due to the short radiative lifetime of polaritons) might appear as a drawback, it allows access to both phase and amplitude of the polariton quantum fluid in a continuous non-destructive manner for as long as the condensate is being replenished by the incoherent excitonic reservoir.

The sample used here is the same as in our previous studies [7],[11]. It was excited by non-resonant subpicosecond pulses (250 fs pulses at 695nm). A high numerical aperture microscope objective collecting the luminescence emitted by the condensate (at ~740nm) ensures a diffraction limited real space resolution. Access to the phase of the polariton quantum fluid is obtained by means of interferometric measurements performed with a modified Michelson interferometer in the mirror-retroreflector configuration. Vortex $2\pi$ phase singularities leave a clear signature on the interference fringes with the form of a forklike dislocation.

Randomly moving vortices cannot be seen in a time averaged experiment but are rather expected to wash out the interference pattern reducing the observed contrast. Surprisingly enough, clear forklike dislocations can be identified in time integrated interferograms yielding the existence of, possibly, pinned quantum vortices. A typical time-integrated interferogram at the output of the interferometer is shown in Fig. 1(a) where the pinning location of a vortex is highlighted by the red circle, centered at the forklike dislocation of the interference pattern. From the interference fringes it is possible to extract the phase of the polariton quantum fluid, which inevitably changes by $2\pi$ as one goes around the vortex core, as shown in Fig. 1(b). To gain access to the dynamics, the interferogram was sent to the entrance slits of a streak camera where it was "cut" into multiple slices with real-space and temporal resolution throughout a tomographic decomposition [12]. This allowed for a full reconstruction of the two-dimensional interferogram from 0 to 155 ps with 3 ps resolution given by the streak camera.

The forklike dislocation observed in the time integrated interferogram is initially not pinned. It exhibits a smooth motion along a well defined path towards the center of the condensate. The most interesting migration dynamics is found within the first 35 ps of the condensate life. The vortex appears close to the side of the excitation spot simultaneously with the formation of the condensate at about 20 ps after the excitation pulse arrival. The subsequent smooth vortex motion is then abruptly interrupted at 55 ps when the vortex encounters a potential trap created by the stationary disorder that acts as a pinning center.

The snap-shot interferograms taken during this time interval are shown in Fig. 2. For visibility purposes the constant component of the 2D interferograms is removed and the color scale of the remaining interference pattern is saturated, giving a very high contrast. At the sample location we studied here, the condensate of exciton-polaritons was linearly polarized with a polarization pinned to one of the crystal axis, suppressing the existence of half-quantum vortices detected at different places on the same sample [13].

Fig. 3a shows the real space path of the vortex as extracted from the interferometric snapshots (shown in Fig. 2). During 35 ps the vortex migrates by about 3 $\mu m$. The velocity of the vortex migration is an inhomogeneous function of time. The vortex velocity profile is shown in Fig. 3(b) with a general trend showing an early period with almost constant velocity followed by a significant deceleration to zero indicating the pinning of the vortex.

The clear observation of the vortex movement is highly surprising having in mind that the interferometric images are the result of an ensemble averaging over millions of independent experiments corresponding to a train of excitation pulses. In the case of stochastic motion of vortices, the averaging would wash out each particular trajectory. Our observation indicates the existence of some preferential paths that vortices follow in thousands of experiments.

In order to better understand the appearance of preferential vortex paths we have performed a numerical modeling using the stochastic generalized Gross-Pitaevskii equation. In our experiment, the non-resonant excitation creates a hot electron-hole plasma that further relaxes to form high-energy excitons [14], which we separate in two subsets: (i) The "inactive" excitons that do not fulfill the required energy and momentum conservation conditions to scatter directly into the condensate state, described by the density $n_I(\mathbf{r},t)$ and (ii) the "active" excitons that do fulfill those conditions and act as the condensate source, described by the density $n_A(\mathbf{r},t)$.

The initial "inactive" exciton concentration $n_I(\mathbf{r},t)$ is created by the external pump $P(\mathbf{r},t)$. These excitons disappear by nonradiative recombination, which is described by the decay rate $\gamma_I$, or they can turn into the "active" excitons at a rate $1/\tau_R$. Those excitons may further relax into the condensate. The dynamics of optically injected excitons is given by a rate equation:

$$\frac{\partial n_I(\mathbf{r},t)}{\partial t} = -\gamma_I n_I(\mathbf{r},t) - \frac{1}{\tau_R} n_I(\mathbf{r},t) + P(\mathbf{r},t) \qquad (1)$$

The "active" exciton population may decay radiatively (with a rate $\gamma_A$) or be scattered to the condensate of exciton-polaritons. This scattering is stimulated by the population of the condensate. The rate equation for the reservoir of "active" excitons reads:

$$\frac{\partial n_A(\mathbf{r},t)}{\partial t} = -\left(\gamma_A + R_R |\psi(\mathbf{r},t)|^2\right) n_A(\mathbf{r},t) + \frac{1}{\tau_R} n_I(\mathbf{r},t) \qquad (2)$$

where $R_R$ is the rate at which excitons scatter into the condensate. $\psi(\mathbf{r},t)$ is the mean scalar polariton field.

The equation for the polariton field that we use here is the well-known Gross-Pitaevskii equation [15], which has previously been used to describe the nonlinear spatial dynamics in various microcavity experiments [11],[12]:

$$i\hbar\frac{\partial \psi(\mathbf{r},t)}{\partial t} = \left\{ -\frac{\hbar^2 \nabla^2}{2m} + g|\psi(\mathbf{r},t)|^2 + V(\mathbf{r}) + g_R n_A(\mathbf{r},t) + \right.$$
$$\left. + g_R n_I(\mathbf{r},t) - \frac{i}{2}\left[\gamma_c - R_R n_A(\mathbf{r},t)\right] \right\} \psi(\mathbf{r},t) \quad (3)$$

The interactions between condensed polaritons in the mean-field approximation are characterized by the constant $g$ whereas interactions between incoherent excitons and polaritons are accounted for by $g_R$. The stationary disorder $V(\mathbf{r})$ has been chosen as a randomly generated Gauss correlated disorder [16]. Polaritons decay from the condensate at a rate $\gamma_c$. Note that it is important to distinguish between "active" and "inactive" excitons, as we did since once stimulated scattering into the condensate takes place, the reservoir of active excitons is depleted. Experimentally we observe a condensate lifetime longer than the polariton lifetime and so we deduce that depleted "active" excitons must be replenished by "inactive" excitons in order to maintain the condensate. The stochastic element in our model is given by the initial condition $\psi(\mathbf{r},0) = \theta(r)/2 \cdot dr$ where θ is a stochastic noise, drawn from a Gauss distribution and characterized by the correlators $\langle \theta^*(r), \theta(r') \rangle = 2\delta(r-r')$, $\langle \theta(r), \theta(r') \rangle = 0$ with $dr$ being the grid spacing. This initial condition implies an average mode occupancy of ½ and random phase, which samples the Wigner distribution of a vacuum state and physically represents the quantum noise of the polariton vacuum.

The relaxation of excitons from the reservoir to the condensate that is triggered by the initial noise, amplifies the density creating small isolated regions with well defined phase. Merging of these regions with random relative phase at the transition to form a state with macroscopic phase coherence leaves initially many topological defects in the phase. This spontaneous vortex generation process is a manifestation of the Kibble-Zurek mechanism [4][5].

Eqs. (1-3) can be solved numerically and each realization of the initial noise term corresponds to the arrival of a different pulse in the experiment. Averaging over multiple realizations of the noise is equivalent to the experimental procedure where the images registered by the streak camera result from averaging over multiple pulses [17]. Different initial conditions result in an altered distribution of vortices. The kinetics of vortices is defined by the interplay between the local disorder potential and the non equilibrium character that induces flows of polaritons. Vortices are dragged by the existent flows resulting in a vortex migration.

A comparison of the evolution of the condensate for different initial conditions yields two distinct kinds of vortices: those that tend to have random paths during the condensate evolution and those that follow specific space-time trajectories. This interesting latter kind of vortices is seen to appear not necessarily in every, but in most of the noise realizations. The vortices which have similar trajectories during the evolution of the condensation tend to survive the averaging and manifest themselves by a clear phase singularity in the time-resolved ensemble averaged interferograms.

The simulations give us direct access to the phase and amplitude of the polariton fluid which we interfere with a reference field resulting in interferograms directly comparable to those experimentally observed. The orange points in Fig. 4(a)-(d), represent the locations of the spontaneously generated vortices for four subsequent instants (40, 50, 60 and 90 ps after the condensate formation) of the condensate evolution averaged over the different experimental realizations. The disorder potential profile is depicted by the contour plots. A comparison of these images reveals two important features: (i) the density of vortices changes as time increases. The modification of the vortex concentration with time can be understood by the possible processes of vortex annihilation (generation) through vortex-antivortex binding (unbinding) or the escape of vortices from the finite-size condensate. The disappearance of random unpinned vortices from the condensate is indicative of the establishment of a long range spatial order. (ii) The preferential vortex locations are in the vicinity of extrema of the disorder because the flow of polaritons in and out of the extrema favors the pinning of a vortex at those locations. Note also the high vortex density regions close to the edges of the condensate. They are indicative of poor spatial phase coherence in the low polariton density regions. Fig. 4(e)-(h) shows the ensemble averaged interferograms for the same four instants of the condensate evolution as in Fig. 4(a)-(d).

The interferograms show forklike dislocations which follow specific trajectories, characterizing a correlated migration of vortices in different experimental realizations. One of the migrating forklike dislocations is indicated by colored circles. The red circle shows its final location as shown in Fig. 4(h). We note that not every pulse contributes a vortex moving along such a trajectory. This is the main reason for the reduced final contrast of the interference fringes. In addition, the regions of high vortex density at the edges of the cloud feature a low contrast of the interference fringes due to the reduced coherence.

In conclusion, a thorough experimental and theoretical investigation of the dynamics of spontaneously created quantum vortices is presented. Quantum vortices are shown to arise from the Kibble-Zurek mechanism. By means of advanced high temporal resolution interferometric techniques, the spontaneous vortices are found to follow well defined space-time trajectories and finally get pinned by the stationary disorder. Our analysis allows us to conclude that although between different pulses certain differences in the vortex trajectories occur, the ensemble averaged interferograms do present the forklike dislocations provided the vortices in different experimental realizations are close enough in space and time. The stable paths are defined by the flows of polaritons in the disorder potential which "drags" along the randomly generated vortices.

We thank Y. Rubo, T. Ostatnicky, M. Toledo Solano and Y. Leger for fruitful discussions. The work was supported by the Swiss National Research Foundation through "NCCR Quantum Photonics".

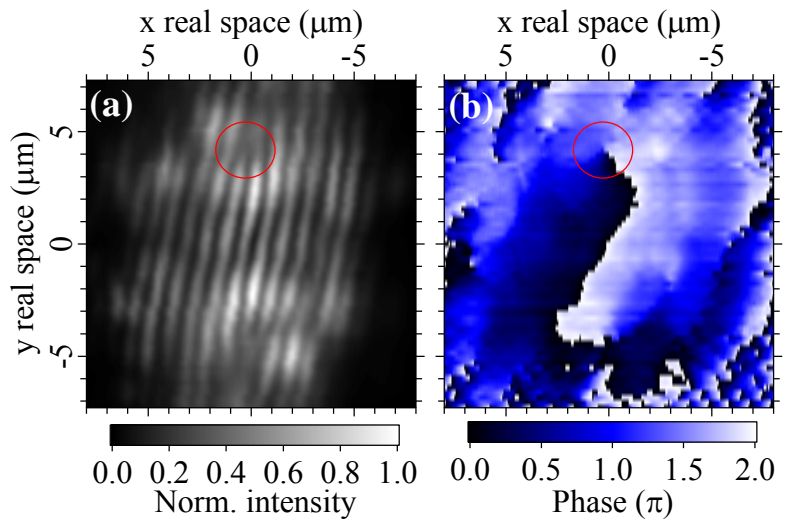

FIG. 1:(a) Time integrated interferogram after the reconstruction of the temporally resolved real space luminescence. The autocorrelation was set to (0,0). The forklike dislocation created by the phase singularity of the vortex is indicated by a red circle and appears in two symmetric sites of the image,

(0.3,4.4) and (-0.3,-4.4), because of the action of the retroreflector. (b) Phase of the polariton fluid with the $2\pi$ vortex signature highlighted with the red circle.

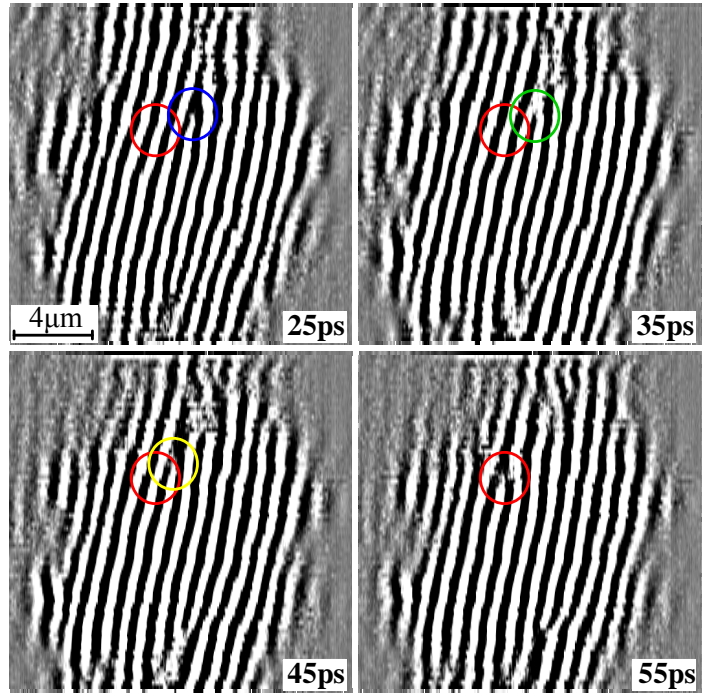

FIG. 2: Dynamics of the migration of the vortex. Each image corresponds to a subsequent time frame (25, 35, 45 and 55 psec). The red circle shows the final vortex location in each time frame whereas the colored circles show the current vortex location. The cw part of the interferograms has been here removed and the contrast is saturated for readability purposes.

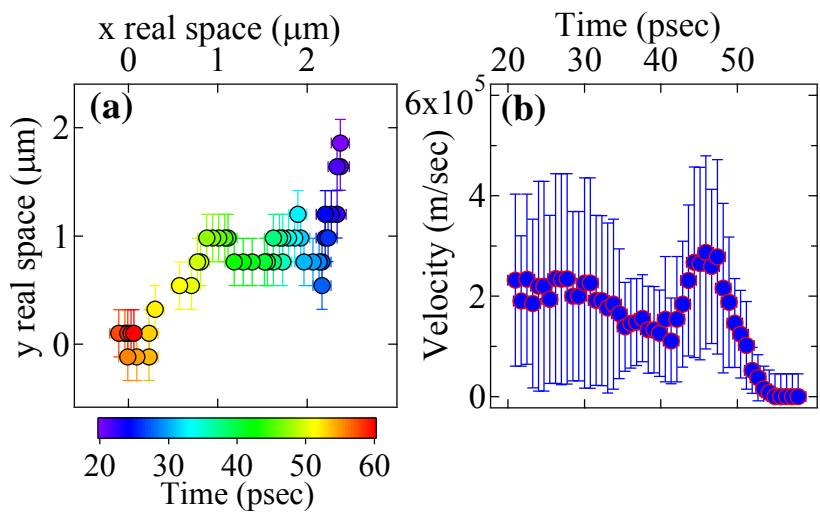

FIG. 3:(a) Real space coordinates of the vortex at different times as denoted by the color of the markers. (b)Velocity of the vortex plotted as a function of time. The large error bars are due to the poor resolution induced by the tomographic technique.

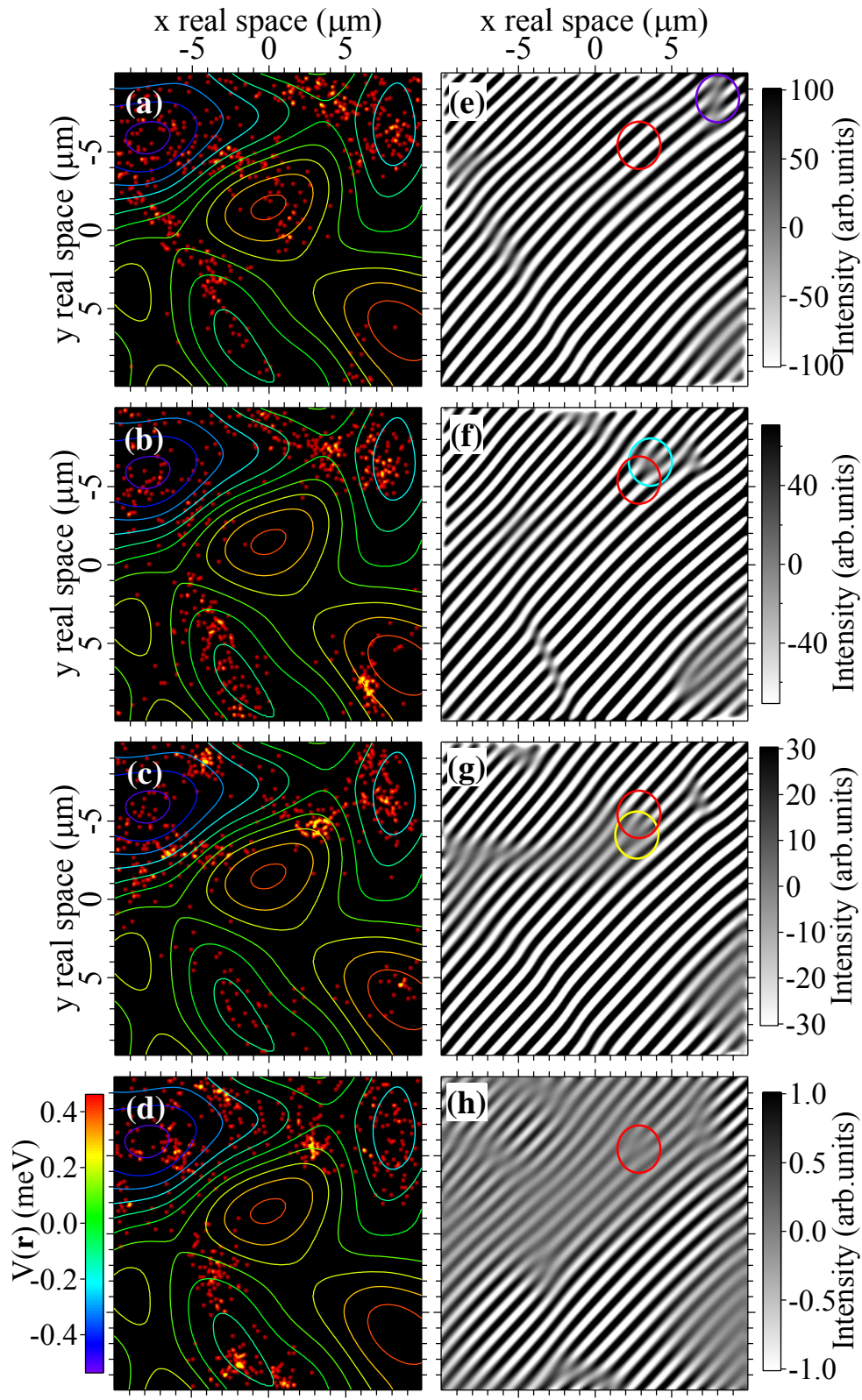

Fig. 4: (a)-(d) Simulation of real space vorticity map for subsequent time frames (40, 50, 60 and 90 ps) where the orange pixels correspond to the appearance of vortices under different initial condition

realizations. The colored contours depict the local disorder potential. (e)-(h) Realization averaged interference patterns for the same time frames as in (a)-(d). Many forklike dislocations migrate in time to different real space locations. The red circle shows the final location of a pinned vortex. The colored circles show the current position of the specific vortex for the different time frames.